\documentclass[a4paper,12pt]{article}

\usepackage{amsmath}
\usepackage[psamsfonts]{amssymb}
\usepackage{rsfs}
\usepackage{bm}

\usepackage{cite}
\usepackage[dvips]{graphicx}
\usepackage{color}

\begin{document} 

\begin{titlepage}

\baselineskip 10pt
\hrule 
\vskip 5pt
\leftline{}
\leftline{Chiba Univ. Preprint
          \hfill   \small \hbox{\bf CHIBA-EP-146}}
\leftline{\hfill   \small \hbox{hep-th/0404252}}
\leftline{\hfill   \small \hbox{September 2004}}
\vskip 5pt
\baselineskip 14pt
\hrule 
\vskip 1.0cm
\centerline{\Large\bf 
Magnetic condensation,     
} 
\vskip 0.3cm
\centerline{\Large\bf  
Abelian dominance, 
}
\vskip 0.3cm
\centerline{\Large\bf  
and instability of Savvidy vacuum 
}
\vskip 0.3cm
\centerline{\large\bf  
}

\vskip 0.5cm

\centerline{{\bf 
Kei-Ichi Kondo$^{\dagger,{1}}$ 
}}  
\vskip 0.5cm
\centerline{\it
${}^{\dagger}$Department of Physics, Faculty of Science, 
Chiba University, Chiba 263-8522, Japan
}
\vskip 1cm

\begin{abstract}
 We show that a certain type of  color magnetic condensation  originating from  magnetic monopole configurations is sufficient to provide the mass for  off-diagonal gluons in the SU(2) Yang-Mills theory under the Cho--Faddeev--Niemi decomposition.  
We point out that the generated gluon mass can cure the instability of the Savvidy vacuum.
In fact, such a novel type of magnetic condensation is shown to occur by calculating the effective potential. 
This enables us to explain the infrared Abelian dominance and monopole dominance by way of a non-Abelian Stokes theorem, which  suggests the dual superconductivity picture of quark confinement.  
Finally, we discuss the implication to the Faddeev-Skyrme model with knot soliton as a low-energy effective theory of Yang-Mills theory.

\end{abstract}


\hrule  
\vskip 0.1cm
${}^1$ 
  E-mail:  {\tt kondok@faculty.chiba-u.jp}

\par 
\par\noindent


\vskip 0.5cm

\newpage
\pagenumbering{roman}




\end{titlepage}


\pagenumbering{arabic}

\baselineskip 14pt
\section{Introduction}

As a promising mechanism of quark confinement, 
dual superconductivity picture of QCD vacuum has been intensively investigated  up to today since the early proposal in 1970s 
\cite{dualsuper}. 
The key ingredient of this picture is the existence of {\it monopole condensation} which causes the dual Meissner effect.  As a result, the color electric flux between quark and antiquark is squeezed to form the string.  Then the non-zero string tension plays the role of the proportional constant of the linear potential realizing quark confinement. 
For this picture to be true,  the monopole condensed vacuum is expected to give more stable vacuum than the perturbative one. 

On the other hand, Savvidy \cite{Savvidy77} has argued based on the general theory of the renormalization group that the  {\it spontaneous generation of color magnetic field} should occur in Yang-Mills theory. However, 
Nielsen and Olesen \cite{NO78} have shown immediately that 
the explicitly calculated effective potential for the magnetic field in the background gauge has a pure imaginary part in addition to the real part which agrees with the prediction of the renormalization group equation. 
In other words, the Savvidy vacuum with magnetic condensation is unstable due to the existence of the {\it tachyon mode}  corresponding to the lowest Landau level realized by the applied external color magnetic field. 
It should be remarked that the off-diagonal gluons were treated as if they were massless in these calculations.  

Now we recall that the infrared (IR) Abelian dominance \cite{tHooft81,EI82} and magnetic monopole dominance in the Maximal Abelian (MA) gauge were confirmed in 1990s by numerical simulations \cite{SY90} where the MA gauge is the same as the background gauge adopted in \cite{NO78}.  
The infrared Abelian dominance \cite{EI82} follows from {\it non-zero mass of off-diagonal gluons}, see \cite{AS99,BCGMP03} for numerical simulations and \cite{Schaden99,DV03,Kondo01,EW03,KondoI} for analytical studies. 
The off-diagonal gluon mass should be understood as the dynamical mass or the spontaneously generated mass, since the original Yang-Mills Lagrangian does not include the mass term. 
If sufficiently large dynamical mass is generated for off-diagonal gluons to compensate for the tachyon mode due to spontaneously generated magnetic field, therefore, the instability might disappear to recover the stability of the Savvidy vacuum. 

The main purpose of this paper is to clarify the physical origin of the dynamical mass for the off-diagonal gluons and thereby to  obtain thorough understanding for the infrared Abelian dominance and monopole dominance. 
In this connection, we discuss    
how the above two pictures (dual superconductivity and magnetic condensation) are compatible with each other in the gauge-independent manner.
This issue is also related to 
how to treat the IR divergence originating from the massless gluons and the tachyon mode corresponding to the lowest Landau level. 
In other words, we examine whether a self-consistent picture can be drawn among  
magnetic condensation, massive off-diagonal gluons, Abelian dominance, monopole dominance and   stability of Savvidy vacuum (or disappearance of tachyon mode). 
\footnote{
Recently, the instability of the Savvidy vacuum
has been re-examined in \cite{Cho03,IM03} from different viewpoints. 
}

In this paper we adopt as  a technical ingredient 
the Cho-Faddeev-Niemi (CFN) decomposition \cite{Cho80,FN98} to extract the topological configurations 
for the  SU(2) Yang-Mills theory in the MA gauge.



\section{Magnetic condensation and massive gluons}
\setcounter{equation}{0}

\subsection{Cho-Faddeev-Niemi decomposition}

We adopt the Cho-Faddeev-Niemi (CFN) decomposition for the non-Abelian gauge field \cite{Cho80,FN98} for extracting the topological configurations explicitly, such as magnetic monopoles (of Wu-Yang type) and multi-instantons (of Witten type).   
By introducing  a three-component vector field $\bm{n}(x)$ with unit length, i.e., 
$\bm{n}(x) \cdot \bm{n}(x) = 1$, in the SU(2) Yang-Mills theory, 
the non-Abelian gauge field $\mathscr{A}_\mu(x)$ is decomposed as
\begin{align}
   \mathscr{A}_\mu(x) 
=\overbrace{ \underbrace{ A_\mu(x) \bm{n}(x)}_{\mathbb{A}_\mu(x)}
 +    \underbrace{ g^{-1}  \partial_\mu \bm{n}(x)  \times \bm{n}(x) }_{\mathbb{B}_\mu(x)}}^{\mathbb{V}_\mu(x)}  
 + \mathbb{X}_\mu(x) ,
\end{align}
where we have used the notation: 
$\mathbb{A}_\mu(x):=A_\mu(x) \bm{n}(x)$,  
$\mathbb{B}_\mu(x):=g^{-1} \partial_\mu \bm{n}(x) \times \bm{n}(x)$
and
$\mathbb{V}_\mu(x):=\mathbb{A}_\mu(x)+\mathbb{B}_\mu(x)$. 
By definition,  
$\mathbb{A}_\mu$ is parallel to $\bm{n}$, while $\mathbb{B}_\mu$ is orthogonal to $\bm{n}$.  We require $\mathbb{X}_\mu$ to be orthogonal to $\bm{n}$,  i.e., 
$\bm{n}(x) \cdot \mathbb{X}_\mu(x)=0$.  
We call $\mathbb{A}_\mu(x)$ the restricted potential, while $\mathbb{X}_\mu(x)$ is called the gauge-covariant potential and 
 $\mathbb{B}_\mu(x)$ can be identified with the non-Abelian magnetic potential as shown shortly. 
In the naive Abelian projection,%
\footnote{
The naive Abelian projection corresponds to adopting the field $\bm{n}(x)$ uniform in spacetime, e.g., $\bm{n}(x)=(0,0,1)$.  Then $\mathbb{V}_\mu(x)$ has the off-diagonal part $V_\mu^a(x)=B_\mu^a(x) (a=1,2)$ and the diagonal part $V_\mu^3(x)=A_\mu(x)$.  See \cite{Cho03} for details. 
}
$\mathbb{A}_\mu(x)$ corresponds to the diagonal component, 
while $\mathbb{X}_\mu(x)$ corresponds to the off-diagonal component, 
apart from the magnetic part $\mathbb{B}_\mu(x)$. 
Accordingly, the non-Abelian field strength $\mathscr{F}_{\mu\nu}(x)$ is decomposed as
\begin{align}
  \mathscr{F}_{\mu\nu} := \partial_\mu \mathscr{A}_\nu - \partial_\nu \mathscr{A}_\mu + g \mathscr{A}_\mu \times \mathscr{A}_\nu
= \mathbb{F}_{\mu\nu} + \mathbb{H}_{\mu\nu} + \hat{D}_\mu \mathbb{X}_\nu - \hat{D}_\nu \mathbb{X}_\mu + g \mathbb{X}_\mu \times \mathbb{X}_\nu , 
\end{align}
where we have introduced the covariant derivative, 
$
  \hat{D}_\mu[\mathbb{V}]  \equiv \hat{D}_\mu  := \partial_\mu   + g \mathbb{V}_\mu \times  ,
$
and defined the two kinds of {\it Abelian} field strength:
\begin{align}
  \mathbb{F}_{\mu\nu} =& F_{\mu\nu} \bm{n}, \quad
F_{\mu\nu} := \partial_\mu A_\nu - \partial_\nu A_\mu ,
\\
  \mathbb{H}_{\mu\nu}  
=& \partial_\mu \mathbb{B}_\nu - \partial_\nu \mathbb{B}_\mu + g \mathbb{B}_\mu \times \mathbb{B}_\nu
=  - g \mathbb{B}_\mu \times \mathbb{B}_\nu 
=  - g^{-1}   (\partial_\mu \bm{n} \times \partial_\nu \bm{n})   .
\end{align}
Here $\mathbb{H}_{\mu\nu}$ is the magnetic field strength which is proportional to $\bm{n}$ and is rewritten in the form
\begin{align}
  \mathbb{H}_{\mu\nu} =& H_{\mu\nu} \bm{n}, \quad
H_{\mu\nu} := - g^{-1} \bm{n} \cdot (\partial_\mu \bm{n} \times \partial_\nu \bm{n})  
= \partial_\mu h_\nu - \partial_\nu h_\mu ,
\end{align}
since $H_{\mu\nu}$ is shown to be locally closed and it can be exact locally with the Abelian magnetic potential $h_\mu$. 
This is because  $H_{\mu\nu}$ represents the color magnetic field generated by magnetic monopoles as follows.
The parameterization of the unit vector
\begin{align}
   \bm{n}(x) 
:= \begin{pmatrix} n^1(x) \cr n^2(x) \cr n^3(x) \end{pmatrix}
:= \begin{pmatrix} \sin \theta(x) \cos \varphi(x) \cr
 \sin \theta(x) \sin \varphi(x) \cr
\cos \theta(x)  
 \end{pmatrix}  ,
\label{paramet}
\end{align}
leads to the expression \cite{KondoII}
\begin{align}
 \bm{n} \cdot (\partial_\mu \bm{n}   \times \partial_\nu \bm{n}) 
 = \sin \theta (\partial_\mu \theta \partial_\nu \varphi
 - \partial_\mu \varphi \partial_\nu \theta) 
 = \sin \theta {\partial(\theta, \varphi) 
 \over \partial(x^\mu,x^\nu)}  ,
\end{align}
where 
${\partial(\theta, \varphi) \over \partial(x^\mu,x^\nu)}$ 
is the Jacobian of the transformation from the coordinates 
$(x^\mu,x^\nu)$ on  
$S^2_{phy}$ to $S^2_{int}$ parameterized
by $(\theta, \varphi)$. 
Then the surface integral of $H_{\mu\nu}$ over a closed surface is quantized in agreement with the Dirac quantization condition:
\begin{align}
 Q_m :=&  \oint_{S^2_{phy}} d\sigma_{jk} 
  g^{-1} \bm{n} \cdot (\partial_j \bm{n}   \times \partial_k \bm{n}) 
 \nonumber\\
=& g^{-1} \oint_{S^2_{phy}} d\sigma_{jk} 
 \sin \theta {\partial(\theta, \varphi)  \over \partial(x^j,x^k)}
 =  g^{-1}  \oint_{S^2_{int}} \sin \theta d\theta  d\varphi 
= {4\pi \over g}  n  ,
\end{align}
where $S^2_{int}$ is a surface of a unit sphere with area $4\pi$. 
Hence
$Q_m$ gives a number of times  $S^2_{int} \simeq SU(2)/U(1)$ is
wrapped by a mapping from  $S^2_{phys} $ to 
$S^2_{int}$. 
This fact is mathematically characterized as the non-trivial second Homotopy class, $\Pi_2(SU(2)/U(1)) = \Pi_2(S^2) = \mathbb{Z}$. 
The Abelian magnetic  potential $h_\mu$ is given by 
$h_\mu=g^{-1}[\cos \theta \partial_\mu \varphi + \partial_\mu \chi]$ with arbitrary $\chi$. 
See \cite{KondoII} for more information. 


The original gauge transformation, 
$
  \delta_\omega \mathscr{A}_\mu  =  D_\mu[\mathscr{A}] \bm{\omega}  ,
$
is replaced by two sets of gauge transformations. 
\footnote{
Note that the way of separating the original gauge transformation is unique as far as two conditions are preserved, i.e., 
$\bm{n}(x) \cdot \bm{n}(x) = 1$ and $\bm{n}(x) \cdot \mathbb{X}_\mu(x)=0$, see \cite{KMS04} for more details.
}

\underline{Gauge transformation I}: (passive or quantum gauge transformation)
\begin{subequations}
\begin{align}
  \delta_\omega \bm{n}  =& 0  ,
\\
 \delta_\omega A_\mu =& \bm{n} \cdot  D_\mu[\mathscr{A}] \bm{\omega} ,
\quad 
  \delta_\omega \mathbb{X}_\mu =     D_\mu[\mathscr{A}] \bm{\omega} - \bm{n}( \bm{n} \cdot  D_\mu[\mathscr{A}] \bm{\omega}) ,
\\ 
 ( \Longrightarrow  & \delta_\omega \mathbb{B}_\mu  =   0 ,
\quad
  \delta_\omega \mathbb{V}_\mu 
=   \bm{n}( \bm{n} \cdot  D_\mu[\mathscr{A}] \bm{\omega})  ). 
\end{align}
\end{subequations}

\underline{Gauge transformation II}: (active or background gauge transformation)
\begin{subequations}
\begin{align}
  \delta_{\omega}' \bm{n}  =& g \bm{n} \times \bm{\omega'}  ,
\\
 \delta_{\omega}' A_\mu =&    \bm{n} \cdot \partial_\mu \bm{\omega'}   ,
\quad 
  \delta_{\omega}' \mathbb{X}_\mu =  g \mathbb{X}_\mu \times \bm{\omega'} ,
\\
(\Longrightarrow  & \delta_{\omega}' \mathbb{B}_\mu  
=   D_\mu[\mathbb{B}] \bm{\omega'} - (\bm{n} \cdot \partial_\mu \bm{\omega'}) \bm{n}  ,
\quad
\delta_{\omega}' \mathbb{V}_\mu =   D_\mu[\mathbb{V}] \bm{\omega'}   ) . 
\end{align}
\end{subequations}
The gauge transformation II does not mix $\mathbb{X}_\mu$ with the Abelian variables $\bm{n}$ and $A_\mu$. 
The transformation law for the field strength can be obtained in the similar way.

\subsection{Magnetic condensation and mass of off-diagonal gluons}

The massiveness of the off-diagonal gluons implies the Abelian dominance \cite{EI82}.  
First, we show that the magnetic condensation defined below leads to the massive off-diagonal gluons at least in the MA gauge.  
Therefore, we argue that the Abelian dominance follows from the magnetic condensation.  In the last stage we argue that this result is also consistent with the monopole dominance and monopole condensation. 

By making use of the  CFN decomposition, the (Euclidean) Yang-Mills Lagrangian is rewritten as 
\begin{align}
  \mathscr{L}_{YM} := \frac{1}{4} \mathscr{F}_{\mu\nu}^2 
=& \frac{1}{4} (\partial_\mu \mathscr{A}_\nu - \partial_\nu \mathscr{A}_\mu + g \mathscr{A}_\mu \times \mathscr{A}_\nu )^2
\nonumber\\
=& \frac{1}{4} (\mathbb{F}_{\mu\nu} + \mathbb{H}_{\mu\nu} + g \mathbb{X}_\mu \times \mathbb{X}_\nu )^2
+ \frac{1}{4} (\hat{D}_\mu \mathbb{X}_\nu - \hat{D}_\nu \mathbb{X}_\mu )^2 , 
\end{align}
where we have used a fact that 
$\mathbb{F}_{\mu\nu}$, $\mathbb{H}_{\mu\nu}$, $g \mathbb{X}_\mu \times \mathbb{X}_\nu$, and hence 
$\mathbb{F}_{\mu\nu} + \mathbb{H}_{\mu\nu} + g \mathbb{X}_\mu \times \mathbb{X}_\nu$ 
are all parallel to 
$\bm{n}$, while $\hat{D}_\mu \mathbb{X}_\nu - \hat{D}_\nu \mathbb{X}_\mu$
 is orthogonal to $\bm{n}$ (which follows from the fact $\bm{n} \cdot \mathbb{X}_\mu=0$). 
By collecting the terms in $\mathbb{X}_\mu$, the Lagrangian reads
\begin{align}
  \mathscr{L}_{YM} 
=& \frac{1}{4} (\mathbb{F}_{\mu\nu} + \mathbb{H}_{\mu\nu})^2
+ \frac{1}{2}  \mathbb{X}_\mu \cdot \mathbb{W}_{\mu\nu} \cdot \mathbb{X}_\nu  
+ \frac{1}{4} (g \mathbb{X}_\mu \times \mathbb{X}_\nu )^2 ,
\end{align}
where we have defined
$\mathbb{X}_\mu \cdot \mathbb{W}_{\mu\nu} \cdot \mathbb{X}_\nu: = X_\mu^A W_{\mu\nu}^{AB} X_\nu^B$ with 
\begin{align}
   W_{\mu\nu}^{AB} 
 :=&    2g \epsilon^{ABC} n^C  [F_{\mu\nu} + H_{\mu\nu}]  
 -(D_\rho[\mathbb{V}] D_\rho[\mathbb{V}])^{AB} \delta_{\mu\nu} 
+ (D_\mu[\mathbb{V}] D_\nu[\mathbb{V}])^{AB} .
\label{defQ}
\end{align}
The second term of $W_{\mu\nu}^{AB}$ is cast into the form,
\begin{align}
 -(D_\rho[\mathbb{V}] D_\rho[\mathbb{V}])^{AB}  
=  -\partial_\rho^2 \delta^{AB} + g^2 [(V_\rho^C)^2 \delta^{AB} - V_\rho^A V_\rho^B] - g \epsilon^{ABC} \partial_\rho V_\rho^C - 2g \epsilon^{ABC} V_\rho^C \partial_\rho ,
\end{align}
where
$V_\rho^A=(\mathbb{A}_\rho)^A+(\mathbb{B}_\rho)^A$ and 
$(V_\rho^A)^2 = \mathbb{V}_\rho \cdot \mathbb{V}_\rho = A_\rho A_\rho + \mathbb{B}_\rho \cdot \mathbb{B}_\rho$.
We focus on the quadratic term in $\mathbb{X}_\mu$ without derivatives,  
\begin{align}
  \frac{1}{2} g^2 [(V_\rho^C)^2 \delta^{AB} - V_\rho^A V_\rho^B] X_\mu^A   X_\mu^B , 
\label{crossterm}
\end{align}
which can be diagonalized to yield the mass term for the off-diagonal gluons. 
Therefore, {\it if the vacuum condensation} 
\begin{align}
 \langle \mathbb{B}_\rho(x) \cdot \mathbb{B}_\rho(x)  \rangle \not= 0 
\end{align}
{\it occurs, then  the off-diagonal gluons acquire the non-zero mass} $M_X$ given by 
\begin{align}
  M_X^2  
=    g^2 \langle \mathbb{B}_\rho(x) \cdot \mathbb{B}_\rho(x)  \rangle 
= \langle (\partial_\rho \bm{n}(x))^2  \rangle  .
\label{mass}
\end{align}
Note that $\mathbb{B}_\mu \cdot \mathbb{B}_\mu=(\partial_\mu \bm{n})^2$ has the gauge invariance I, while it does not have the gauge invariance II, but the minimum value of 
the spacetime average of $\mathbb{B}_\mu(x) \cdot \mathbb{B}_\mu(x)$, i.e.,
$min_{\omega'}\int_{\Omega} d^4x \mathbb{B}_\mu^{\omega'}(x) \cdot \mathbb{B}_\mu^{\omega'}(x)/\int_{\Omega} d^4x$ 
can have a definite non-zero value \cite{GSZ01}, see also \cite{Kondo01}.
Hereafter, we call the condensation 
$\langle \mathbb{B}_\rho \cdot \mathbb{B}_\rho  \rangle$ 
{\it magnetic condensation} for convenience. 
The magnetic potential $\mathbb{B}_\mu$ is related to the magnetic monopole in the sense explained above. 
If such magnetic condensation takes place at all, it could be related to the (magnetic) monopole condensation. We postpone the relationship between magnetic condensation and (magnetic) monopole condensation to a subsequent paper \cite{KMS04}.%
\footnote{
In this paper we do not consider the condensation 
$
 \langle \mathbb{V}_\rho \cdot \mathbb{V}_\rho  \rangle
= \langle A_\rho A_\rho  \rangle 
+ \langle \mathbb{B}_\rho \cdot \mathbb{B}_\rho  \rangle 
$
or
$
\langle (A_\rho+h_\rho)^2  \rangle 
$, 
paying special attention to the magnetic potential $\mathbb{B}_\rho$ or  $h_\rho$ alone. 
This is partly because the electric (photon) condensation 
$ \langle A_\rho A_\rho  \rangle$ 
will be incompatible with the residual U(1) invariance, and partly because we can not obtain the final closed expression for the effective potential including the electric U(1) potential $A_\rho$. 
}

\subsection{Two types of magnetic condensation}

Next, we discuss how such magnetic condensation can occur actually in gluodynamics. 
At first sight, such a situation looks very difficult to be realized, since the original Yang-Mills Lagrangian does not include the quartic polynomials in the field $\mathbb{B}_\mu$ of the type $(\mathbb{B}_\rho \cdot \mathbb{B}_\rho)^2$, although the other type $(\mathbb{B}_\mu \times \mathbb{B}_\nu)^2$ is contained as a part of $\mathbb{H}_{\mu\nu}^2$ in the tree Lagrangian.%

We restrict our arguments to the Euclidean space. 
Then a mathematical identity for the $SU(2)$ gauge potential,  
yields the lower bound on $\mathbb{B}_\mu \cdot \mathbb{B}_\mu$,
$
  (\mathbb{B}_\mu \cdot \mathbb{B}_\mu)^2 - (\mathbb{B}_\mu \times \mathbb{B}_\nu) \cdot (\mathbb{B}_\mu \times \mathbb{B}_\nu) = (\mathbb{B}_\mu \cdot \mathbb{B}_\nu) (\mathbb{B}_\mu \cdot \mathbb{B}_\nu) \ge 0, 
$
i.e.,
\begin{align}
   g \mathbb{B}_\mu \cdot \mathbb{B}_\mu    \ge \sqrt{(g\mathbb{B}_\mu \times \mathbb{B}_\nu)^2} \equiv \|\mathbb{H}\| , 
\end{align}
where 
$
 \|\mathbb{H}\|  :=   \sqrt{\mathbb{H}_{\mu\nu} \cdot \mathbb{H}_{\mu\nu}}
$
and
$  
\mathbb{H}_{\mu\nu} \cdot \mathbb{H}_{\mu\nu} = (\partial_\mu \mathbb{B}_\nu - \partial_\nu \mathbb{B}_\mu + g \mathbb{B}_\mu \times \mathbb{B}_\nu)^2 
=  (g\mathbb{B}_\mu \times \mathbb{B}_\nu)^2 .
$
Therefore, the magnetic condensation in question 
$\langle g \mathbb{B}_\rho \cdot \mathbb{B}_\rho  \rangle \not=0$
follows from the other magnetic condensation 
$\langle \|\mathbb{H}\|  \rangle 
= \langle \sqrt{(g\mathbb{B}_\mu \times \mathbb{B}_\nu)^2}  \rangle 
\not= 0$, since 
\begin{align}
  \langle g \mathbb{B}_\rho \cdot \mathbb{B}_\rho  \rangle 
\ge  \langle \sqrt{(g\mathbb{B}_\mu \times \mathbb{B}_\nu)^2}  \rangle
\equiv \langle \|\mathbb{H}\|  \rangle \not= 0 .
\label{ineq}
\end{align}
Thus, 
$\langle g \mathbb{B}_\rho \cdot \mathbb{B}_\rho  \rangle$ and $\langle \|\mathbb{H}\|  \rangle$ are not independent. 
To obtain the definite value of $\langle g \mathbb{B}_\rho \cdot \mathbb{B}_\rho  \rangle$, we need to know $\langle (\mathbb{B}_\mu \cdot \mathbb{B}_\nu)^2  \rangle$ in addition to $\langle \|\mathbb{H}\|  \rangle$ which will be calculated below.

\subsection{Gauge invariance and BRST quantization}

The naive generating functional is given by ($\mathscr{L}_{J}$: source term)
\begin{align}
 Z[J]
=  \int \mathcal{D}A_\mu \mathcal{D}\mathbb{X}_\mu    \mathcal{D}\bm{n}\delta(\bm{n} \cdot \bm{n}-1)  \delta(\bm{n} \cdot \mathbb{X}_\mu) \delta(D_\mu[\mathbb{V}] \mathbb{X}_\mu)
 e^{ - \int_{x} \mathscr{L}_{YM} - \int_{x} \mathscr{L}_{J}  } . 
\end{align}
The measure $\mathcal{D}A_\mu \mathcal{D}\mathbb{X}_\mu    \mathcal{D}\bm{n}$ is shown to be invariant under the I, II gauge transformations.%
\footnote{
 For the integral measure for the CFN variables to be completely equivalent to the standard measure of the gauge theory, it is necessary \cite{Shabanov99} to add a determinant 
$\Delta_{S}:=\det \left|{\delta {\chi} \over \delta \bm{n}} \right|_{{\chi}=0}$.  However, the determinant does not influence the one-loop result, see   \cite{Shabanov99,Gies01}, and it is neglected in the following. 

} 
Two conditions 
$\bm{n} \cdot \bm{n}-1=0$ and $\bm{n} \cdot \mathbb{X}_\mu=0$ are also invariant under the gauge transformations I and II. 
The additional condition 
$
 \bm{\chi} := D_\mu[\mathbb{V}] \mathbb{X}_\mu = 0 
$ 
is covariant under the gauge transformation II, i.e.,
$\delta_{\omega'} (D_\mu[\mathbb{V}] \mathbb{X}_\mu) = g (D_\mu[\mathbb{V}] \mathbb{X}_\mu) \times \bm{\omega'}$, while 
 it is neither invariant nor covariant under the gauge transformation I.  
Then it can be identified with the gauge fixing (GF) condition for the off-diagonal part perpendicular to $\bm{n}$ (generalized differential MA gauge) according to \cite{Shabanov99}, since  
  $\bm{n} \cdot D_\mu[\mathbb{V}] \mathbb{X}_\mu=0$. 
In fact, it is obtained by minimizing $\int d^4x {1 \over 2} \mathbb{X}_\mu \cdot \mathbb{X}_\mu$ w.r.t. the  gauge transformation I:
\begin{align}
   0 = \delta_\omega \int_{x} {1 \over 2} \mathbb{X}_\mu \cdot \mathbb{X}_\mu 
= - \int_{x} \bm{\omega} \cdot  D_\mu[\mathbb{V}] \mathbb{X}_\mu  .
\end{align}

We can define two sets of BRST transformations $\bm{\delta},\bm{\delta}'$ corresponding to gauge transformations I, II by introducing the Nakanishi-Lautrup (NL) auxiliary fields $\mathbb{N}, \mathbb{N}'$ and the Faddeev-Popov (FP) ghost $\mathbb{C}, \mathbb{C}'$ and antighost fields $\bar{\mathbb{C}},\bar{\mathbb{C}}'$, respectively.  The BRST transformations defined in this way are shown to be nilpotent $\bm{\delta}\bm{\delta}=0=\bm{\delta}'\bm{\delta}'$.
We can obtain the two FP ghost terms $\mathscr{L}_{FP}$ and $\mathscr{L}_{FP}'$ for the MA GF condition, 
\begin{align}
  Z[J] =& \int \mathcal{D}\mathbb{C} \mathcal{D}\bar{\mathbb{C}} \mathcal{D}\mathbb{N}
\mathcal{D}\mathbb{C}' \mathcal{D}\bar{\mathbb{C}}' \mathcal{D}\mathbb{N}'
\int \mathcal{D}\mathbb{X}_\mu 
\int \mathcal{D}\bm{n}\delta(\bm{n} \cdot \bm{n}-1) 
 \delta(\bm{n} \cdot \mathbb{X}_\mu) 
\int \mathcal{D}A_\mu   
\nonumber\\& \times
  e^{ - \int_{x} [\mathscr{L}_{YM}+\mathscr{L}_{GF+FP}+\mathscr{L}_{GF+FP}' ]  - \int_{x}  \mathscr{L}_{J} }  , 
\end{align}
where (apart from the GF + FP term for the parallel part)
\begin{align}
  \mathscr{L}_{GF+FP} =& 
-i \bm{\delta}(\bar{\mathbb{C}} \cdot D_\mu[\mathbb{V}]\mathbb{X}_\mu)
\nonumber\\ 
=&  \mathbb{N} \cdot D_\mu[\mathbb{V}]\mathbb{X}_\mu 
+ i \bar{\mathbb{C}} \cdot D_\mu[\mathbb{V}]D_\mu[\mathscr{A}] \mathbb{C}
+  g^2 (i\bar{\mathbb{C}} \cdot (\bm{n} \times \mathbb{X}_\mu))
((\bm{n} \times \mathbb{X}_\mu) \cdot \mathbb{C}) ,
\nonumber\\
  \mathscr{L}_{GF+FP}' =& -i \bm{\delta}'(\bar{\mathbb{C}}' \cdot D_\mu[\mathbb{V}]\mathbb{X}_\mu)
=  \mathbb{N}' \cdot D_\mu[\mathbb{V}]\mathbb{X}_\mu 
+ ig \bar{\mathbb{C}}' \cdot (D_\mu[\mathbb{V}]\mathbb{X}_\mu \times \mathbb{C}') . 
\label{GF+FP2}
\end{align}

In this framework, it is not difficult to show \cite{KMS04} that the effective action written in terms of $\bm{n}$ and $A_\mu$  is obtained in the one-loop level by integrating out $\mathbb{X}_\mu$ and the other fields  as 
\begin{subequations}
\begin{align}
 Z[J] =& \int \mathcal{D}\bm{n}\delta(\bm{n} \cdot \bm{n}-1) 
\int \mathcal{D}A_\mu   
  e^{ - S_{EFF}   }  , 
\\
S_{EFF} =& \int_{x} \frac{1}{4} [\mathbb{F}_{\mu\nu} + \mathbb{H}_{\mu\nu} ]^2
+  \frac{1}{2}\ln \det \{ (\mathbb{Q}){}_{\mu\nu} \} 
- \ln \det \{ -D_\mu[\mathbb{V}]D_\mu[\mathbb{V}] \}  
\nonumber\\&
+   \frac{1}{2} \ln \det \{ D_\mu[\mathbb{V}] (\mathbb{Q}^{-1}){}_{\mu\nu} D_\nu[\mathbb{V}] \} + \cdots ,
\\
 [\mathbb{Q}]_{\mu\nu}^{AB} 
 :=&  -  (D_\rho[\mathbb{V}] D_\rho[\mathbb{V}])^{AB} \delta_{\mu\nu}  + g \epsilon^{ABC} n^C K_{\mu\nu}   ,
\quad 
 K_{\mu\nu} := 
 2(F_{\mu\nu} +  H_{\mu\nu} ) .
\label{defQ3}
\end{align}
\end{subequations}
It is possible \cite{KMS04} to show in the similar way to  \cite{KondoI,Freire02} that the coupling constant in this framework of the Yang-Mills theory under the CFN decomposition obeys exactly the same beta function as the original Yang-Mills theory, although this issue has already been discussed based on the momentum shell integration of the Wilsonian renormalization group \cite{Gies01}.



\subsection{Calculation of magnetic condensation via effective potential}

Now we proceed to  estimate the logarithmic determinants to obtain the effective potential. 
However, only in the pure magnetic case, i.e., $K_{0i}=0, K_{ij}\not=0$ ($F_{\mu\nu}=0, H_{\mu\nu}\not=0$), we can obtain a closed form for the effective potential of  the (spacetime-independent constant) field $K_{\mu\nu}$.

In case of massless gluons, the eigenvalue $\lambda_n$ of the operator $(\mathbb{Q}){}_{\mu\nu}$ is given by $2g\|\mathbb{H}\|(n+3/2), 2g\|\mathbb{H}\|(n-1/2), 2g\|\mathbb{H}\|(n+1/2), 2g\|\mathbb{H}\|(n+1/2)$ with $n=0,1,2, \cdots$,
corresponding to two transverse physical modes and two unphysical  (longitudinal and scalar) modes respectively.  The contribution from two unphysical gluon modes are cancelled by that from the ghost and antighost, since the operator $-D_\mu[\mathbb{V}]D_\mu[\mathbb{V}]$ has exactly the same eigenvalues.  
Only the two physical modes contribute to the mode sum $\sum_{n}  e^{-\lambda_{n} \tau}$ in the heat kernel method as
\begin{align}
     \sum_{n=0}^{\infty} e^{-2g\|\mathbb{H}\|(n+3/2)\tau} 
+  \sum_{n=0}^{\infty} e^{-2g\|\mathbb{H}\|(n-1/2)\tau} 
= {e^{-3g\|\mathbb{H}\| \tau} \over 1-e^{-2g\|\mathbb{H}\| \tau}}
+  {e^{ g\|\mathbb{H}\| \tau} \over 1-e^{-2g\|\mathbb{H}\| \tau}},
\label{level}
\end{align}
and hence the logarithmic determinants are calculated  as
\begin{align}
 & \frac{1}{2}\ln \det \{ (\mathbb{Q}){}_{\mu\nu} \} 
- \ln \det \{ -D_\mu[\mathbb{V}]D_\mu[\mathbb{V}] \}  
+   \frac{1}{2} \ln \det \{ D_\mu[\mathbb{V}] (\mathbb{Q}^{-1}){}_{\mu\nu} D_\nu[\mathbb{V}] \} 
\nonumber\\
 =& - \int d^4x {1 \over (4\pi)^2} \lim_{s \rightarrow 0} {d \over ds} {\mu^{2s} \over \Gamma(s)} \int_{0}^{\infty} d\tau \tau^{s-2} \  2g\|\mathbb{H}\| 
 \left[ {e^{-3g\|\mathbb{H}\| \tau} \over 1-e^{-2g\|\mathbb{H}\| \tau}} 
+ {e^{g\|\mathbb{H}\| \tau} \over 1-e^{-2g\|\mathbb{H}\| \tau}} \right] .
\label{det1}
\end{align}
Then the effective potential $V(\|\mathbb{H}\|)$ for    
$\|\mathbb{H}\|:=(\mathbb{H}_{\mu\nu}^2)^{1/2}$ is obtained as
\footnote{
This one-loop calculation agrees with the result \cite{Cho03}.
An attempt going beyond the one-loop  will be given in a subsequent paper \cite{KMS04}. 
The result (\ref{effpot}) is apparently the same as the old result obtained by Nielsen and Olesen \cite{NO78} in which  it is assumed that  
the external magnetic field $H_{12}$ has the specific direction in color space and the magnitude is  uniform in space-time from the beginning. 
In the present derivation using the CFN decomposition, 
the direction $\bm{n}(x)$ of the color magnetic field 
$\mathbb{H}_{\mu\nu}(x)=H_{\mu\nu}(x) \bm{n}(x)$ 
can be chosen arbitrary at every spacetime  point $x$, and the Lorentz and color rotation symmetry is not broken. 
Moreover, this formalism enables us to specify the physical origin of magnetic condensation as arising from the magnetic monopole 
through the relation, 
$
H_{\mu\nu}(x) := - g^{-1} \bm{n}(x) \cdot (\partial_\mu \bm{n}(x) \times \partial_\nu \bm{n}(x))
$. 
}
\begin{align}
  V(\|\mathbb{H}\|) = {1 \over 4}\|\mathbb{H}\|^2 \left[ 1 + {b_0 \over 16\pi^2}g^2 \left( \ln {g\|\mathbb{H}\| \over \mu^2} -c \right) \right] 
+ \text{pure imaginary part} 
,
\label{effpot}
\end{align}
where $b_0={22 \over 3}$. 
The origin of the pure imaginary part in (\ref{effpot}) is  the last term  in (\ref{level}) and (\ref{det1}), corresponding to the lowest Landau level $n=0$, which does not decrease for large $\tau$ leading to the infrared divergence at $\tau=\infty$, whereas the ultraviolet divergence at $\tau=0$ can be handled by isolating the $s=0$ pole {\it a la} dimensional regularization.
The real part of the effective potential has a minimum at $\|\mathbb{H}\|=\|\mathbb{H}\|_0$ away from the origin. 
Neglecting the pure imaginary part, therefore, the magnetic condensation occurs at the minimum of the effective potential, 
\begin{align}
  \langle \|\mathbb{H}\| \rangle = \|\mathbb{H}\|_0 := \bar{\mu}^2 e^{1} \exp \left[ -{16\pi^2 \over b_0 \bar{g}^2}\right] .
\end{align}

\subsection{Recovering the stability of the Savvidy vacuum}

The appearance of the pure imaginary part due to a tachyon mode is easily understood as follows. 
The formal Gaussian integration over the off-diagonal gluons leads to 
\begin{align}
  \int \mathcal{D}\mathbb{X}_\mu    e^{ -\frac{1}{2} \mathbb{X}_\mu \mathbb{Q} \mathbb{X}_\nu } 
= {1 \over \sqrt{\det \mathbb{Q}}} = e^{-\frac{1}{2} \ln \det \mathbb{Q} } ,
\end{align}
where the determinant is given as the product of all the eigenvalues,  
$
  \det \mathbb{Q} = \prod_{n=0}^{\infty}  \lambda_{n} .
$
A tachyon mode $\lambda_0<0$ makes the $\det \mathbb{Q}$ negative, since the other eigenvalues are positive. 
Hence, the logarithm produces the pure imaginary part. 

Once the dynamical mass generation occurs through the magnetic condensation, the eigenvalues are shifted upward by $M_X^2$, i.e, $\lambda \rightarrow \lambda + M_X^2$.   Especially, the lowest tachyon mode $-g \|\mathbb{H}\|<0$ is modified to $-g \|\mathbb{H}\|+M_X^2$. 
If $M_X^2$ is large enough so that $M_X^2>g \|\mathbb{H}\|$, the tachyon mode is removed, i.e., $-g \|\mathbb{H}\|+M_X^2>0$, and all the eigenvalues of $\mathbb{Q}$ become positive.  
The effective potential can be calculated by summing up the modes with positive eigenvalues alone as in the massless case.
Thus the infrared divergence disappears and the pure imaginary part is eliminated in the effective potential, if the ratio $r :=M_X^2/(g\|\mathbb{H}\|)$ is greater than one, $r > 1$. 

Even when the off-diagonal gluons are massive, we can see that 
the effective potential $V(\|\mathbb{H}\|)$ still has a nontrivial minimum in $\|\mathbb{H}\|$, although the location $\|\mathbb{H}\|'_0$ of the minimum could change, $\|\mathbb{H}\|'_0\not=\|\mathbb{H}\|_0$. 
  This is because the existence of the nontrivial minimum originates from the asymptotic freedom, i.e., positivity of the coefficient $b_0=22/3$ of the quantum correction part $V(\|\mathbb{H}\|)-\frac{1}{4}\|\mathbb{H}\|^2$ in the one-loop  effective potential $V(\|\mathbb{H}\|)$ where $-b_0$ agrees with the first coefficient of the $\beta$-function, $\beta(g)={-b_0 \over 16\pi^2} g^3 + O(g^5)$.  
In the massive case, it is shown that  $b_0$ is modified into   
$
 b'_0 = -8[2\zeta(-1,(3+r)/2)+r] \ge b_0
$, 
depending only on the ratio $r$, 
where $\zeta(z,\lambda)$ is the generalized Riemann zeta function. 
Since $b'_0$ is positive for arbitrary value of $r  \ge 0$, the effective potential $V(\|\mathbb{H}\|)$ still has a nontrivial minimum at $\|\mathbb{H}\|=\|\mathbb{H}\|'_0$ without the pure imaginary part for $r>1$.
(Of course, the constant $c$ is also changed and hence the location of the minimum $\|\mathbb{H}\|'_0$ could change, but it does not affect the existence of the minimum.)
In fact, the bound $r \ge 1$, i.e., 
$M_X^2\ge g\|\mathbb{H}\|'_0$ follows from the identification (\ref{mass}) and the inequality (\ref{ineq}), 
$M_X^2=g^2\|\mathbb{B}_\mu^2\|\ge g\|\mathbb{H}\|'_0$, provided that the expectation value $\langle \|\mathbb{H}\|  \rangle$ is given by the minimum $\|\mathbb{H}\|'_0$ of the one-loop effective potential for massive gluons. 
 (A possible zero mode will be discussed in \cite{KMS04}.)

The mass generation for  off-diagonal gluons due to magnetic condensation modifies  
 the energy mode,
\begin{align}
  \sqrt{k_{\parallel}^2 -g\|\mathbb{H}\|'_0}  \rightarrow \sqrt{k_{\parallel}^2 + M_X^2-g\|\mathbb{H}\|'_0} \ge \sqrt{k_{\parallel}^2  } .
\end{align}
Thus, the tachyon mode is removed and  the instability of the Savvidy vacuum no longer exists.

\subsection{Another view through the change of variables}

\begin{figure}[htbp]
\begin{center}
\includegraphics[height=6cm]{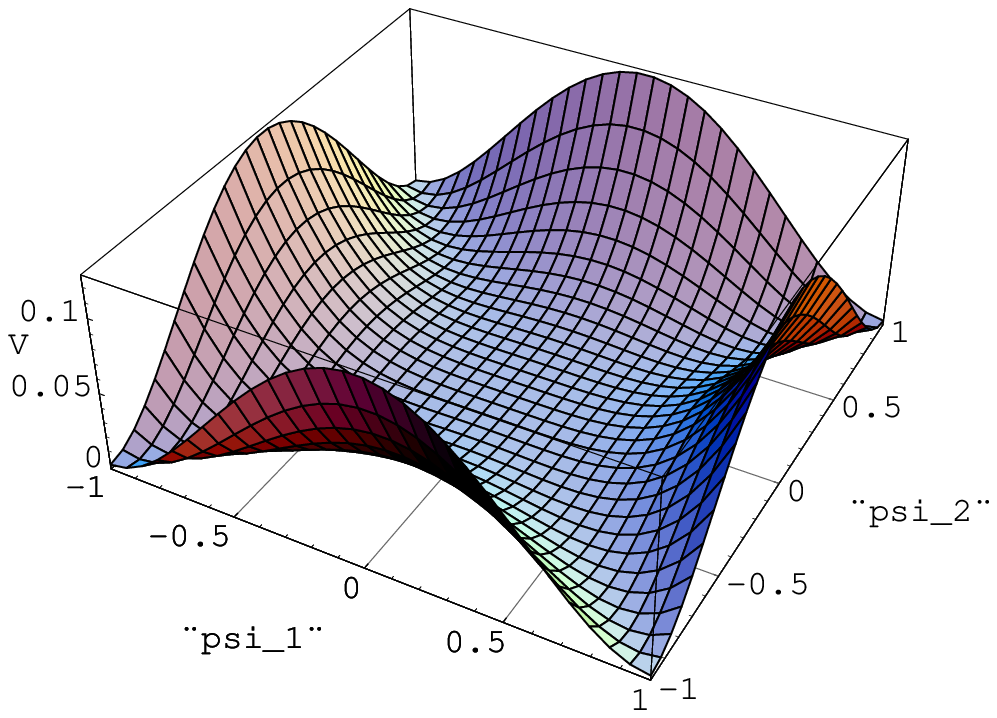}
\includegraphics[height=6cm]{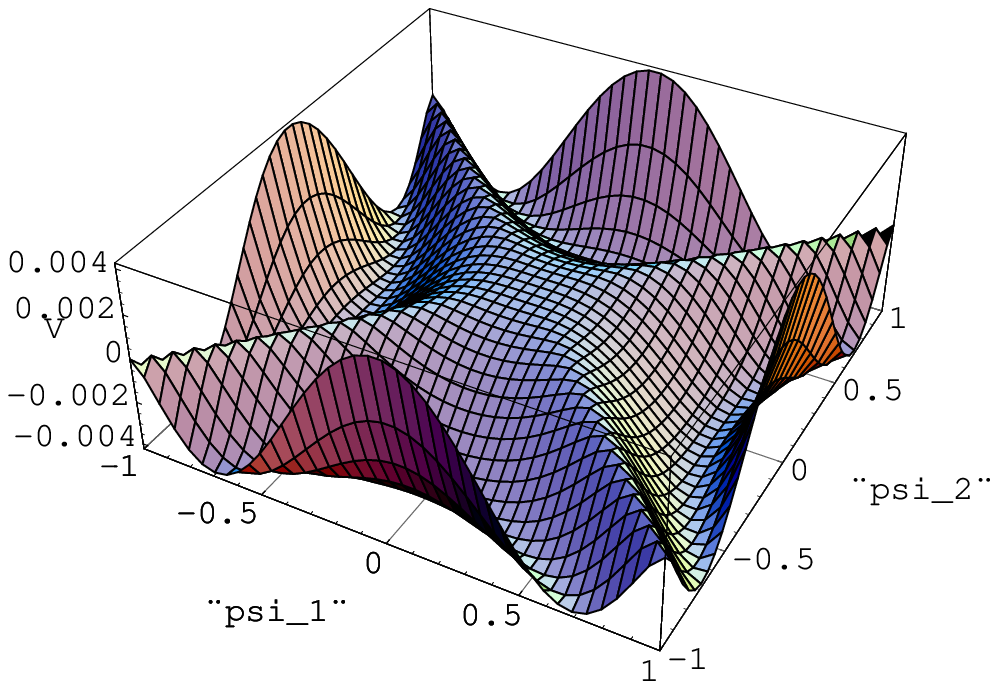}
\caption{\small Effective potential $V$ as functions of 
$|\psi_1|=\frac{1}{2}(|X|+|Y|)$ and $|\psi_2|=\frac{1}{2}(|X|-|Y|)$.
(a) Classical case $V(\|\mathbb{H}\|)=\frac{1}{4}\|\mathbb{H}\|^2$: all the minima are on the two lines $X=0$ or $Y=0$, (b) Quantum case (one-loop level): all the minima lie in the four hyperbolae 
$XY=\pm \sqrt{2} \langle \|\mathbb{H}\| \rangle/g$
forming the flat directions.}
\label{fig:potential}
\end{center}
\end{figure}

Another view is obtained for the effective potential of the magnetic component as shown below. 
We apply the Faddeev-Niemi decomposition \cite{FN02,Freyhult02} to the magnetic potential $\mathbb{B}_\mu(x)$ 
where we introduce the two complex scalar fields $\psi_1, \psi_2 \in \mathbb{C}$ and a tweibein $e_\mu^a(x) (a=1,2)$ with 
$ e_\mu^a(x) e_\mu^b(x) = \delta^{ab}$ to parameterize the plane of a four-dimensional space in which the two vector fields $B_\mu^1$ and $B_\mu^2$ lie:  
\begin{align}
   \sqrt{2} B_\mu^{+} :=& B_\mu^1 + i B_\mu^2 
= i \psi_1 \bm{e}_\mu + i \psi_2 \bm{e}_\mu^* ,  
\end{align}
where   
$
  \bm{e}_\mu = {1 \over \sqrt{2}} (e_\mu^1 + i e_\mu^2 )  
$
is the complex combination satisfying
$\bm{e}_\mu \cdot  \bm{e}_\mu = 0$,  
$\bm{e}_\mu \cdot \bm{e}_\mu^* = 1$  
$(\bm{e}_\mu^* \cdot \bm{e}_\mu^* = 0 )$. 
By making use of this decomposition, two magnetic variables are rewritten as 
\begin{align}
 H^2/g^2 = (\epsilon^{ab} B_\mu^a B_\nu^b)^2
=& {1 \over 2}[|\psi_1|^2-|\psi_2|^2]^2  
= {1 \over 2}  X^2 Y^2   ,
\\
B_\mu^a B_\mu^a    
=& |\psi_1|^2+|\psi_2|^2 
= {1 \over 2} (X^2+Y^2) ,
\end{align}
where we have defined 
$
 X := |\psi_1|+|\psi_2|, \quad Y:= |\psi_1|-|\psi_2| .
$
The effective potential $V(\|\mathbb{H}\|)$ is drawn as function  of new variables $X,Y$ or $|\psi_1|,|\psi_2|$ in Fig.~\ref{fig:potential}.
The circle (cylinder) intersects with the four hyperbolae if the radius exceeds the minimal value, 
$\frac{1}{2}(X^2+Y^2) \ge \sqrt{X^2Y^2}=|XY|$, i.e., 
$g^2 \langle B_\mu^a B_\mu^a \rangle \ge \sqrt{2} g \langle \|\mathbb{H}\| \rangle$, see Fig.~\ref{fig:potential}. 
Thus the result of the one-loop calculation is consistent with the non-vanishing vacuum condensation 
$\langle g \mathbb{B}_\rho \cdot \mathbb{B}_\rho  \rangle \not=0$.  

This mechanism of providing the mass for the gauge boson (off-diagonal gluons)   is very similar to the Higgs mechanics of providing the mass for the gauge particles through the non-vanishing vacuum expectation value (VEV) of the scalar field 
$\phi_0=\langle \phi \rangle \not=0$
 which is obtained as the location $\phi_0$ of the absolute minimum of the Higgs potential $V(\phi)$.  However, the tree potential has no non-trivial minima and the potential has non-trivial minima only through the radiative corrections  based on the Coleman--Weinberg mechanism \cite{CW73}.


\subsection{Wilson loop and Abelian dominance}

In order to show quark confinement, we need to calculate the Wilson loop average
$\langle W_C[\mathscr{A}] \rangle$.  
The non-Abelian Wilson loop operator $W_C[\mathscr{A}]$ for a closed loop $C$ is defined as the trace of the path-ordered product $\mathscr{P}$ of the exponential of the line integral of the gauge potential along the closed loop $C$:
\begin{align}
  W_C[\mathscr{A}] 
  := {\rm tr} \left[ \mathscr{P} \exp \left\{ ig \oint_{C} dx^\mu \mathscr{A}_\mu(x) \right\} \right]  .
\end{align} 
A version of the non-Abelian Stokes theorem \cite{DP89,KondoIV}
 for SU(2)  
 and \cite{KT99} for SU(N) 
allows us to rewrite the Wilson loop into the surface integral over the surface $S$ with the boundary $C$:
\begin{align}
  W_C[\mathscr{A}]  =& \int d\mu_S(\bm{n}) \exp \left\{ i {J \over 2} \int_{S: \partial S=C} d\sigma^{\mu\nu} G_{\mu\nu} \right\} , 
  \\
  G_{\mu\nu}(x) =& \partial_\mu [\bm{n}(x) \cdot \mathscr{A}_\nu(x)] -  \partial_\nu [\bm{n}(x) \cdot \mathscr{A}_\mu(x)]   
  - {1 \over g} \bm{n}(x) \cdot [\partial_\mu \bm{n}(x) \times \partial_\nu \bm{n}(x) ] ,
\end{align}
where we have used the adjoint orbit representation \cite{KondoIV}
\begin{align}
   \bm{n}(x) = n^A(x) T^A = U^\dagger(x) T^3 U(x) ,
\label{aorep}
\end{align}
and $d\mu_S(\bm{n})$ is the product of the invariant measure on $SU(2)/U(1)$, 
$
  d\mu(\bm{n}(x)) = {2J+1 \over 4\pi} \delta(\bm{n}(x) \cdot \bm{n}(x)-1) d^3 \bm{n}(x) ,  
$
over the surface $S$ with $J$ characterizing the representation of SU(2) group, $J=1/2, 1, 3/2, \cdots $. 
Under the gauge transformation, $\bm{n}$ transforms in  the adjoint representation,
$
     \bm{n}(x) \rightarrow U^\dagger(x) \bm{n}(x)U(x)  .
$
Then $G_{\mu\nu}$ has the manifestly gauge invariant form, 
$
G_{\mu\nu}(x) =  \bm{n}(x) \cdot \mathscr{F}_{\mu\nu}(x)  
  - {1 \over g} \bm{n}(x) \cdot [D_\mu \bm{n}(x) \times D_\nu \bm{n}(x) ] ,
$
with the covariant derivative $D_\mu :=\partial_\mu + g \mathscr{A}_\mu \times $. 
In the CFN decomposition,  $G_{\mu\nu}$ is identified with
\begin{align}
  G_{\mu\nu} = F_{\mu\nu} + H_{\mu\nu} ,
\end{align}
and the Wilson loop is rewritten in terms of the Abelian components 
composed of $A_\mu$ and $\bm{n}$.
If there are no string singularities in $A_\mu:=\bm{n} \cdot \mathscr{A}_\mu$ (say the photon part), then the magnetic monopole current defined by
$k_\mu := \partial^\nu {}^*G_{\mu\nu}$ reduces to
$k_\mu := \partial^\nu {}^*H_{\mu\nu}$
where 
$k_\mu$ obeys the 
topological conservation law $\partial^\mu k_\mu=0$ and is not the Noether current.
It is obvious that 
$G_{\mu\nu}$ has the same form as the 't Hooft--Polyakov tensor under the identification
$
  n^A(x) \leftrightarrow \hat{\phi}^A(x):=\phi^A(x)/|\phi(x)| .
$
Thus the Wilson loop is the probe of the magnetic monopole. 

These results suggest that the massive fields $\mathbb{X}_\mu$ are redundant for large loop $C$ (decouple in the long distance) and the  infrared Abelian dominance in the Wilson loop average is justified to hold. Thus we obtain a consistent scenario of Abelian dominance and monopole dominance for quark confinement \cite{KMS04}.



\subsection{Faddeev-Skyrme model and knot soliton}

It is important to remark that the {\it off-diagonal gluon condensation} \cite{Kondo01} in the present framework  
\begin{align}
\langle  \mathbb{X}_\rho \cdot \mathbb{X}_\rho  \rangle
= 2\Lambda^2 \not= 0 ,
\end{align}
yields the {\it mass term} for the field $B_\mu$:
\begin{align}
  g^2 \Lambda^2  \mathbb{B}_\mu \cdot \mathbb{B}_\mu 
=    \Lambda^2  (\partial_\mu \bm{n})^2  , 
\end{align}
in view of (\ref{crossterm}). 
Therefore, the off-diagonal gluon condensation yields the Faddeev-Skyrme model with knot solitons,
\footnote{
The Faddeev-Skyrme model with knot soliton has been derived in the static limit 
from the SU(2) Yang-Mills theory in the MA gauge, if 
$
  \langle (A_\mu^1)^2 + (A_\mu^2)^2 \rangle = \Lambda^2 \not= 0 ,
$
based on intricate arguments. 
We do not need the procedure for taking the static limit. 
}
\begin{align}
  \mathscr{L} =   \Lambda^2 (\partial_\mu \bm{n})^2 + \frac{1}{2}  [\bm{n} \cdot (\partial_\mu \bm{n} \times \partial_\nu \bm{n})]^2   .
\end{align}
However, the systematic derivation of the off-diagonal gluon condensation is still lacking.%
\footnote{After submitting this paper for publication, an analytic derivation of the off-diagonal mass has appeared in the modified MA gauge \cite{KondoII,KT99}, see \cite{Dudaletal04}.  It is desirable to extend the calculation to the present framework. 
} 
The claim $M_X^2  
=    g^2 \langle \mathbb{B}_\mu(x) \cdot \mathbb{B}_\mu(x)  \rangle 
= \langle (\partial_\mu \bm{n}(x))^2  \rangle \not=0$
is also very suggestive for the existence of the Faddeev-Skyrme model.

\section{Conclusion and discussion}

We have applied 
the Cho-Faddeev-Niemi decomposition to the SU(2) Yang-Mills theory and performed the BRST quantization.  First, we have pointed out that the magnetic condensation, 
$\langle g \mathbb{B}_\rho \cdot \mathbb{B}_\rho  \rangle \not=0$, 
 can provide the mass for the off-diagonal gluons.  Therefore the existence of the magnetic condensation enables us to explain the infrared Abelian dominance which has been confirmed by numerical simulations in the last decade as a key concept for understanding the dual superconductivity. 

 Second, the magnetic condensation in question can occur if the other magnetic condensation takes place, i.e., 
$\langle \sqrt{(g\mathbb{B}_\mu \times \mathbb{B}_\nu)^2}  \rangle \not=0$,  
based on a simple mathematical identity.

Third, we have explicitly calculated the effective potential of the color magnetic field 
$\|\mathbb{H}\|:=\sqrt{(g\mathbb{B}_\mu \times \mathbb{B}_\nu)^2}$ 
in one-loop level.  
The resulting effective potential has a non-zero absolute minimum supporting the magnetic condensation
$\langle g \mathbb{B}_\rho \cdot \mathbb{B}_\rho  \rangle \not=0$. 
We have applied the change of variables to visualize the potential just as the Higgs potential of the scalar field, apart from the Savvidy instability.

Note that the magnetic condensation is not exactly equivalent to the monopole condensation. Yet, two condensations are intimately connected to each other. 
In fact, a scenario for the off-diagonal gluons to acquire their mass for realizing the monopole condensation has  already been demonstrated in \cite{Kondo00}   
where the physical origin of the off-diagonal gluon mass was not specified. 
Therefore, {\it monopole condensation as the origin of dual superconductivity follows from the magnetic condensation} 
$\langle g \mathbb{B}_\rho \cdot \mathbb{B}_\rho  \rangle \not=0$, 
if we regard the resulting massiveness of the off-diagonal gluons. 
As the magnetic condensation is sufficient to provide the mass for off-diagonal gluons, the magnetic condensation and monopole condensation can give a self-consistent picture for supporting the dual superconductivity. 

However, the  existence of the pure imaginary part in the effective potential $V(\|\mathbb{H}\|)$ was a signal of the Savvidy instability which was discovered by Nielsen and Olesen \cite{NO78}.
In this paper we have shown that the mass generation for off-diagonal gluons due to magnetic condensation modifies the calculation of the effective potential and eliminates the pure imaginary part corresponding to the tachyon mode.
Thus, the mass generation for off-diagonal gluons due to magnetic condensation 
 recover the stability of the Savvidy vacuum.  

To go beyond the one-loop calculation presented in this paper, we can introduce the antisymmetric auxiliary tensor field  $\mathbb{B}_{\mu\nu}$ as in \cite{KondoI} and include its condensation as an independent degrees of freedom to search for the true vacuum.  This is also necessary to discuss the confining string \cite{Kondo00,Kondo03}. 
Extension of the magnetic condensation to SU(3) is also indispensable to discuss the realistic world.  
These issues will be tackled in a subsequent paper \cite{KMS04}.

\section*{Acknowledgments}
The author would like to thank Atsushi Nakamura, Takeharu Murakami and Toru Shinohara for helpful discussions. 
This work is supported by 
Grant-in-Aid for Scientific Research (C)14540243 from Japan Society for the Promotion of Science (JSPS), 
and in part by Grant-in-Aid for Scientific Research on Priority Areas (B)13135203 from
The Ministry of Education, Culture, Sports, Science and Technology (MEXT).


\end{document}